\documentclass[traditabstract]{aa} 
\usepackage{graphicx}
\usepackage{txfonts}
\usepackage{natbib}
\bibpunct{(}{)}{;}{a}{}{,} 
\begin{document}
   \title{On inelastic hydrogen atom collisions in stellar atmospheres}

   \author{P. S. Barklem\inst{1}
          \and
          A. K. Belyaev\inst{1,2}
          \and
          M. Guitou\inst{3}
          \and
          N. Feautrier\inst{4}
          \and
          F. X. Gad\'ea\inst{5}
          \and
          A. Spielfiedel\inst{4}
          }

   \institute{Department of Physics and Astronomy, Uppsala University, Box 515 S-75120 Uppsala, Sweden 
         \and
             Department of Theoretical Physics, Herzen University, St.\ Petersburg 191186, Russia
         \and 
             Universit{\'e} Paris-Est, Laboratoire Mod{\'e}lisation et Simulation Multi-Echelle, MSME UMR 8208 CNRS,5 Bd Descartes, 77454 Marne-la-Vall{\'e}e, France
         \and
             LERMA and UMR 8112 of CNRS, Observatoire de Paris-Meudon, 92195 Meudon Cedex, France
         \and
             Laboratoire de Chimie et Physique Quantique, UMR 5626 du CNRS, IRSAMC,
             Universit\'e Paul Sabatier, 118 rte de Narbonne, F-31062, Toulouse, France                
             }

   \date{Received 18 Feb 2011 ; accepted ----}

 
  \abstract
   {The influence of inelastic hydrogen atom collisions on non-LTE spectral line formation has been, and remains to be, a significant source of uncertainty for stellar abundance analyses, due to the difficulty in obtaining accurate data for low-energy atomic collisions either experimentally or theoretically.  For lack of a better alternative, the classical ``Drawin formula'' is often used.   Over recent decades, our understanding of these collisions has improved markedly, predominantly through a number of detailed quantum mechanical calculations.  In this paper, the Drawin formula is compared with the quantum mechanical calculations both in terms of the underlying physics and the resulting rate coefficients.  It is shown that the Drawin formula does not contain the essential physics behind direct excitation by H atom collisions, the important physical mechanism being quantum mechanical in character.  Quantitatively, the Drawin formula compares poorly with the results of the available quantum mechanical calculations, usually significantly overestimating the collision rates by amounts that vary markedly between transitions.}


   \maketitle
%

\section{Introduction}

Solution of the non-local thermodynamic equilibrium (non-LTE) radiative transfer problem in stellar atmospheres requires detailed and complete knowledge of the radiative and collisional processes that affect the statistical equilibrium of a given atomic species of interest.  A difficulty regarding the collisional processes is to determine which, among the almost endless possibilities in a stellar atmosphere, are important.  In an early study of the formation of the Na D lines in the solar spectrum, \citet{1955MNRAS.115..256P} considered the two obvious candidates for the case of the solar atmosphere: inelastic collisions with electrons and hydrogen atoms in their ground state.  The importance of electron collisions in many environments is well known, arising from the fact that electrons are always the most abundant charged particle and have a much higher thermal velocity than atoms, leading directly to a higher collision rate.  

Hydrogen atoms are the most abundant perturber by far; however, as pointed out by Plaskett, given that in thermal conditions atoms are much slower than electrons, simple considerations based on the Massey criterion \citep{1949RPPh...12..248M} indicate H atom collisions are nearly adiabatic and thus should have much smaller inelastic cross sections than electron collisions at the relevant thermal velocities \citep[see][for a detailed discussion]{1993PhST...47..186L}.   Plaskett considered experimental estimates for quenching of Na by He and methane, which suggested deexcitation cross sections ranging from non-detectable for He to a small value for methane.  Assuming that quenching by H might be similar to that for methane, Plaskett estimated the inelastic H atom collision rate would be of the same order of magnitude as that for electrons, despite $n(\mathrm{H})/n(\mathrm{e}) \sim 10^4$.  As indicated by Plaskett, if one made the more reasonable assumption that H would have a similar cross section to He, then the rate would probably be negligible.   

Based on this rough expectation, and probably also because of the almost total lack of any data for atom-atom collisions at low energy due to the increased difficulties in calculating and measuring such cross sections compared to the case of electrons \citep[e.g.][]{1955VA......1..277M}, most early non-LTE studies of cool stars during the 1960's and 1970's neglected H atom collisions.  While some authors mentioned the possible need to include such processes \citep[e.g.][]{1962ApJ...136..906V, 1969ApJ...156..695A}, the first to make an attempt at including them quantitatively was the study of solar Na lines by \citet{1975A&A....38..289G}.  That study used the classical formula from \citet{1969ZPhy..225..483D}, though details on how these formula were extended from the case of collisions between like atoms to collisions between unlike atoms are not given.  Gehren found the H atom collisions to be relatively unimportant, less than half of the electron collision rates.  

A later study of Li line formation by \citet{1984A&A...130..319S} used a similar approach, though in this case the generalisation of the formula of Drawin to inelastic H atom collisions was now given.  The resulting formula is often referred to in stellar astrophysics as the ``Drawin formula''.  The key development of this study was that it considered the case of a low-metallicity halo dwarf star, where $n(\mathrm{H})/n(\mathrm{e}) \sim 10^6$.  They found that if H atom collisions were described to order-of-magnitude accuracy by this formula, then indeed the large ratio $n(\mathrm{H})/n(\mathrm{e})$ would be sufficient to offset the lower efficiency and frequency of H atom collisions relative to electrons, and inelastic collisions with H atoms would be the dominant collisional process in cool metal-poor dwarfs.  Thus, the study of Steenbock and Holweger restimulated the idea that H atom collisions might be important in cool stellar atmospheres, and since that study the question has been: does the Drawin formula really provide order-of-magnitude estimates of these processes?  

Recent theoretical and experimental studies have put us in a position to answer this question.   In this paper we will examine the question from two points of view.  First, is there any reasonable expectation that the Drawin formula can provide reasonable results given our present understanding of the physics involved?  This is discussed in Sect.~\ref{sect:phys}.  Second, how do the results from the Drawin formula compare with those from quantum mechanical calculations?  This comparison is made in Sect.~\ref{sect:comp}.  Finally, the main conclusions are summarised.

\section{Physics of inelastic hydrogen collisions}
\label{sect:phys}

In this section we discuss the physics behind the classical Drawin formula.  We then compare with quantum theory and the physics of inelastic H atom collisions as revealed by experiment and detailed quantum scattering calculations based on quantum-chemistry descriptions of the relevant quasi-molecules.

\subsection{The classical Drawin formula}

The Drawin formula is the result of a number of modifications and extensions of the classical formula for ionization of atoms by electron impact due to \citet{thomson1912}.  In Thomson's theory, the bound electron in the target atom is considered as a stationary free classical electron.  The Coulomb interaction of an incident classical electron with the atomic electron is then considered and Thomson calculates the impact parameter for which the deflection of the incident electron results in an energy transfer between electrons corresponding precisely to the ionization potential.  All collisions inside this impact parameter will have a larger energy transfer and thus lead to ionization and so the cross section can be easily calculated.  \citet{1952ZNatA...7..432E} introduced a factor to account for equivalent electrons in the outer shell of the atom, which is usually also applied.

As pointed out by \citet[Sect.4]{1968RvMP...40..564R}, Thomson's theory is remarkable for being able to make a number of qualitative predictions regarding ionization cross sections and their behaviour with impact energy.  However, quantitatively it provides only an order-of-magnitude estimate.  For example, \citet{1962amp..conf..375S} compared with experimental results for H, He, Ne and Ar, and found the classical theory overestimates by a factor of about 5 at low energy.  Seaton explained this from the fact that in quantum mechanical treatments the ionization probability rarely approaches unity even at small impact parameters, while the classical theory assumes the probability is unity for all impact parameters smaller than that leading to a classical energy transfer equal to the ionization potential.  It was also observed that the comparison differed markedly for Ne, a difference that is explained by quantum mechanical properties of the atom,  and cannot be explained by classical theory.  

\begin{figure}
\includegraphics[width=90mm]{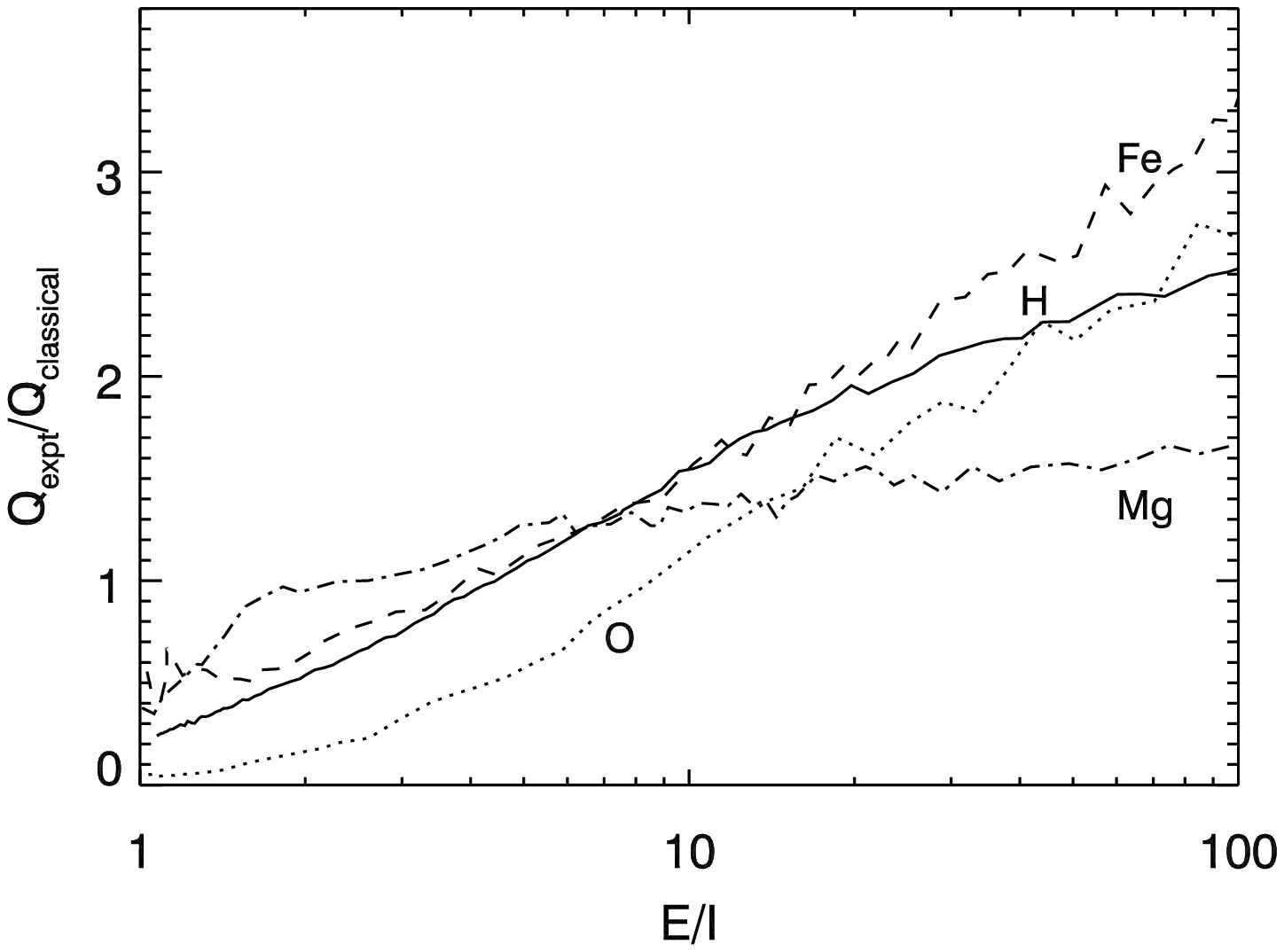}
\caption{Comparison of the classical \citet{thomson1912} cross section for ionization of H, O, Mg and Fe atoms by electron impact with experimental results of \citet{1987JPhB...20.3501S, 1995JPhB...28.1321T, 1992JPhB...25.1051M,1993JPhB...26.2393S}, respectively.  The plot shows the ratio of the experimental values with respect to the theoretical cross section, plotted against the collision energy $E$ scaled by the ionization energy $I$. The errors in the experimental values are in all cases typically 5\% or less, with an additional uncertainty of around 7--14\% in the absolute scale.}
\label{fig:thomson}
\end{figure}

In Fig.~\ref{fig:thomson} we complement Seaton's comparisons with new comparisons with modern experimental data from the Queen's University of Belfast laboratory for the cases of electron impact ionization of ground state hydrogen, oxygen, magnesium and iron \citep[][respectively]{1987JPhB...20.3501S, 1995JPhB...28.1321T, 1992JPhB...25.1051M,1993JPhB...26.2393S}.  These experimental data are in good agreement with other experiments, cover a wide range of collision energies including those near threshold and the absolute uncertainty is expected to be less than 20\% (see caption to Fig.~\ref{fig:thomson}).  The general features of the comparison are very similar to those seen in Seaton's comparison, i.e. the classical formula overestimates the cross sections by around a factor of 5 at low energy.  In this comparison, O is seen to be particularly discrepant, differing by a factor of 20 near threshold.  Thus, quantitively, the Thomson formula is able to provide only a rough order-of-magnitude estimate of the ionization cross section.  Given the difficulties involved in quantum calculations  compared to the classical approach, considerable efforts were made to improve the classical models, predominantly through inclusion of the initial velocity of the atomic electron, the effects of the target nucleus, and exchange effects between electrons.  These efforts were met with mixed success, and there is an extensive literature on the subject \citep[see][for some reviews]{burgess&percival,1968RvMP...40..564R,1970ARA&A...8..329B,1978PhR....35..305B}.

\citet{1968ZPhy..211..404D,1969ZPhy..225..483D, 1973PhLA...43..333D} extended the Thomson theory to the case of ionization and excitation in collisions between like atoms, which was later generalised by \citet{1984A&A...130..319S} to the case of H atom collisions with other atoms.   The derivation has been redone in considerable detail by \citet{1993PhST...47..186L}, correcting an error in the cross section expressions due to confusion of centre-of-mass and laboratory frames; an error that was pointed out earlier by \citet{1972ZPhy..252..435F}.  It is worth noting that if this error was carried through one would expect considerable errors in the rate coefficients due to the incorrect threshold.  However, the equations for the rate coefficients generalised by \citet{1984A&A...130..319S} seem to be correct, with the final expression for the rate coefficient derived by \citet{1993PhST...47..186L} differing from that of Steenbock and Holweger by an essentially negligible factor $m_\mathrm{A}/(m_\mathrm{A}+m_\mathrm{H})$, of order unity.   The resulting expression for the rate coefficient for an inelastic H atom collision with atom A, $\mathrm{A}_{l} + \mathrm{H}  \rightarrow \mathrm{A}_{u} + \mathrm{H}$, where ${l}$ and ${u}$ denote lower and upper states respectively, as presented by Lambert is:
\begin{equation}
\langle \sigma v \rangle = 16 \pi a_0^2 \left( \frac{E_1^\mathrm{H}}{E_{ul}^A} \right) f_{lu} \frac{m_\mathrm{e}(m_\mathrm{A}+m_\mathrm{H})}{(m_\mathrm{H}+m_\mathrm{e})m_\mathrm{H}} \left( \frac{2kT}{\pi \mu} \right)^{1/2} \Psi (w_{ul})
\label{eq:drawin}
\end{equation}
where $f_{lu}$ is the oscillator strength for the transition ${l} \rightarrow {u}$, $E_1^\mathrm{H}$ is the ionization potential of H, $E^\mathrm{A}_{ul} = E^\mathrm{A}_{u} - E^\mathrm{A}_{l}$, $w_{ul} = E^\mathrm{A}_{ul}/kT$, and $\Psi$ is a function arising from the integration over the Maxwell velocity distribution, which for $kT \sim E^\mathrm{A}_{ul}$ is:
\begin{equation}
\Psi(w_{ul}) = \mathrm{exp} (-w_{ul}) \left( 1 + \frac{2}{w_{ul}} \right).
\end{equation}
All other quantities have their usual meaning.

The main physical assumption in the generalisation of the Thomson theory to ionization in collisions between like atoms, is that for a given collision energy above threshold, the efficiency of the energy transfer in an atom-atom collision is the same as in the electron-atom collision.   This amounts to neglect of the nucleus of the perturbing atom, and thus the model in essence is the classical interaction of the impact atom electron with the stationary free classical atomic electron, allowing for the increased mass of the perturber.   Drawin introduces an additional mass factor ($m_\mathrm{A}/m_\mathrm{H}$) without justification, which has been discussed by \citet{1972ZPhy..252..435F} and \citet{1993PhST...47..186L}.    While Drawin compared with experiments for ionization in collisions between light atoms and between molecules and found agreement within a factor of two, Fleischmann \& Dehmel made comparisons and found only order-of-magnitude agreement becoming worse for heavier atoms.   In the case of collisions between atoms of similar mass, Fleischmann \& Dehmel found only agreement within a factor of 100.   Comparisons with more recent experimental data do not change this conclusion \citep{1991JChPh..95.5738K}.

Classical descriptions of atomic processes are less well suited to excitation than to ionization, due to the generally smaller energy transfer and the difficulty in choosing a classical final energy state band.   In (\citealt{1969ZPhy..225..483D}; \citealp[see also][]{1973PhLA...43..333D}) the ionization formula is extended to the case of excitation by analogy with the ionization formula; that is, the ionization threshold energy is replaced by the excitation threshold energy, and the oscillator strength is inserted into the formula.  This amounts to a choice of the final energy band to be all energy transfers greater than the excitation threshold, modified by the oscillator strength.  The inclusion of the oscillator strength is almost certainly motivated by analogy with the Bethe approximation for inelastic collisions with electrons (see below).  This approximation is valid only for optically allowed transitions and at high collision energies where the momentum transfer is small.  The oscillator strength is introduced in some versions of the ionization formula \citep[e.g.][]{1969ZPhy..225..483D} including the one adopted by Steenbock and Holweger, but not in others \citep{1968ZPhy..211..404D,1973PhLA...43..333D}.      

\subsection{Quantum collision theory}

In certain cases, quantum scattering theory leads to a dependence of the excitation 
(and ionization) cross section on the (generalized) oscillator strength. 
This arises within the first Born approximation \citep[e.g.][]{1962amp..conf..375S, LandauLifshitz,BransdenJoachain}, which is valid for high-energy collisions and is obtained within perturbation theory. 
For electron-atom collisions, the unperturbed 
wave functions are taken in the form of atomic initial/final wave functions $\phi^\mathrm{at}_{l(u)}({\bf r})$ 
(${\bf r}= \{{\bf r_\alpha}\}$ denoting a set of coordinates of atomic electrons labeled by $\alpha$)
multiplied by plane wave functions $(2\pi)^{-3/2}\exp(\i{\bf k}_{l(u)} {\bf R})$
of the impact particle (${\bf R}$ being the impact electron coordinate) 
and the perturbation operator 
is the interaction potential $U$
between an impact electron and the target atom, which consists of Coulomb interactions with the nucleus $U_\mathrm{en}$ 
and the atomic electrons $U_\mathrm{ee}$: $U=U_\mathrm{en}+U_\mathrm{ee}$. 
The direct scattering probability amplitude for excitation of the atom $l\to u$ is
\begin{equation}
   A^\mathrm{B1} = -\frac{1}{2\pi} \langle \phi^\mathrm{at}_{u}({\bf r}) | \exp(\imath{\bf q R}) U | \phi^\mathrm{at}_{l}({\bf r})\rangle 
                                                                      \, , \label{eq:1Born}
\end{equation}
where ${\bf q}= {\bf k}_{l}-{\bf k}_{u}$ is the momentum transfer, 
and the integration is taken over both ${\bf r}$ and ${\bf R}$. 
Integration over the impact electron coordinate ${\bf R}$, considering the Coulomb interactions and the orthogonality of the atomic wave functions,
removes the interaction with the nucleus and leaves only the interactions between the incident and atomic electrons 
\begin{equation}
   A^\mathrm{B1} = -\frac{2}{q^2} \langle \phi^\mathrm{at}_{u}({\bf r}) | \sum_\alpha \exp(\imath{\bf q r_\alpha}) | 
            \phi^\mathrm{at}_{l}({\bf r})\rangle_{\bf r} 
                                                                      \, .  \label{eq:1Born2}
\end{equation}
The cross section is proportional to the square of the scattering amplitude in Eq.~\ref{eq:1Born2}, and finally the cross section can be shown to be proportional to the quantity \citep[e.g.][]{BransdenJoachain}
\begin{equation}
\mathcal{F}_{lu}=\frac{2(E_{u}-E_{l})}{q^2}
       |\langle \phi^\mathrm{at}_{u}({\bf r}) | \sum_\alpha \exp(\imath{\bf q r_\alpha}) | \phi^\mathrm{at}_{l}({\bf r})\rangle_{\bf r}|^2 \, , \label{eq:genosc}
\end{equation}
which is called the generalized oscillator strength for reasons that will become apparent.
For optically allowed transitions, expansion of the exponential in Eq.~\ref{eq:genosc} can be restricted to the dipole term leading to a relation between the excitation cross section and both the transition dipole moment  
and the corresponding optical oscillator strength $f_{lu}$; this relation is known as 
the Bethe approximation \citep{1930AnP...397..325B,1962amp..conf..375S,LandauLifshitz,BransdenJoachain}, and leads to
\begin{equation}
\left( A^\mathrm{B1} \right)^2 = \frac{2}{q^2 (E_{u} - E_{l})} f_{lu} \,.
\end{equation}
\citet{1962amp..conf..375S} provides a discussion of the physical interpretation of this relationship between the collisional and radiative process. 
For optically forbidden transitions, the higher-order expansion 
terms resulting from Eq.~\ref{eq:genosc} must be considered thus yielding nonzero excitation/ionization cross sections. 
Thus, the first Born approximation forms a bridge between the quantum approach and the classical approach of Drawin, both being 
based on the Coulomb interactions of an impact charged particle with atomic electrons and 
resulting in relations between the excitation (ionization) cross section and the oscillator strength.

The Born approximation is applicable for high-energy electron-atom and electron-molecule collisions,
but is valid also for charged heavy particle (ion) -- atom/molecule collisions.
For low-energy collisions, which are the main interest of this paper, the Born approximation is no longer valid 
and gives rather poor results (further explaining the poor agreement between 
the classical calculations and the experimental data at low energies, see Fig. 1).

At low collision energies, quantum atomic collision theory uses a different approach, 
in particular, the standard adiabatic (Born-Oppenheimer) approach, which is based on 
the separation of the nuclear kinetic energy operator and the electronic fixed-nuclei Hamiltonian 
in the total Hamiltonian, finally leading to the molecular-state representation for electronic 
(fixed-nuclei) wave functions and nonadiabatic nuclear dynamics 
\citep[e.g.][]{MottMassey,1962amp..conf..550B,1982PhR....90..299M,NikitinUmansky}. 
Total wave functions of the entire collisional system are expressed in terms of products of 
electronic molecular-state wave functions $\psi^\mathrm{mol}_{l(u)}({\bf r};{\bf R})$ 
and nuclear wave functions, e.g., radial nuclear wave functions 
$F_{{l(u)}}(R)/R$ multiplied by angular nuclear wave functions; ${\bf R}$ being nuclear coordinates. 
This full quantum treatment requires calculations of quantum-chemical data (molecular potentials 
and nonadiabatic couplings) and integration of coupled-channel scattering equations. 
The nuclear motion induces nonadiabatic transitions between molecular states. 
In the asymptotic region, where the internuclear distance is large enough, molecular orbitals are determined by 
corresponding atomic orbitals, but at short and intermediate distances, where nonadiabatic transitions mainly 
take place, molecular orbitals are usually represented by mixtures of quite different atomic orbitals, 
for example, covalent and ionic configurations.

Perturbation theory can also be applied to 
low-energy atomic collisions under certain conditions. This can be done in the adiabatic or diabatic representations. 
In both representations the unperturbed functions are molecular-state electronic wave functions, 
adiabatic or diabatic ones, and perturbation operators appear from the nuclear kinetic energy 
operator. 
In the adiabatic representation, the transition probability amplitude reads \citep{2007EPJD...44..497B}
\begin{eqnarray}
    A^\mathrm{ad}  &=& \frac 12 \int_0^\infty \langle \psi^\mathrm{mol}_{l} | \frac{\partial}{\partial R} | \psi^\mathrm{mol}_{u} \rangle_{\bf r} \nonumber \\
    && 
            \times \left[F_{l}^\mathrm{ad}(R) \frac{d F_{u}^\mathrm{ad}(R)}{dR} - F_{u}^\mathrm{ad}(R) \frac{d F_{l}^\mathrm{ad}(R)}{dR} \right] dR
                                                                     \, , \label{eq:A-adiab}
\end{eqnarray}
where 
$\langle \psi^\mathrm{mol}_{l} | {\partial}/{\partial R} | \psi^\mathrm{mol}_{u} \rangle_{\bf r}$ is the nonadiabatic coupling, 
which is responsible for nonadiabatic transitions between adiabatic molecular states $l$ and $u$ 
due to the nuclear motion; see~\citet{2007EPJD...44..497B} for details. 
It is clearly seen from Eq.~\ref{eq:A-adiab} that inelastic transitions in low-energy atomic collisions 
are determined by the nuclear kinetic energy operator 
(i.e. $-\hbar^2 \nabla^2_\mathbf{R} /2\mu$), 
by the mixing of electronic molecular-state wave functions expressed by nonadiabatic couplings, 
and by nuclear wave functions, which substantially deviate from plane waves. 
In the diabatic representation, the corresponding amplitude has the form \citep{NikitinUmansky}
\begin{equation}
    A^\mathrm{diab} =\int_0^\infty 
            F_{l}^\mathrm{diab}(R) H_{lu} F_{u}^\mathrm{diab}(R)dR
                                                                     \, , \label{eq:A-diab}
\end{equation}
where H$_{lu}$ is an off-diagonal matrix element calculated with diabatic electronic molecular-state wave functions
$^\mathrm{diab}\psi^\mathrm{mol}_{{l(u)}}$, 
which are obtained from adiabatic quantum-chemical data by the diabatisation procedure under the condition 
$\langle ^\mathrm{diab}\psi^\mathrm{mol}_{l} | {\partial}/{\partial R} | ^\mathrm{diab}\psi^\mathrm{mol}_{u} \rangle_{\bf r} = 0$ \citep{1969PhRv..179..111S}. 
Again, the basis for nonadiabatic transitions is the nuclear kinetic energy operator, the molecular-state 
wave functions and nuclear wave functions, which should be calculated in this case with diabatic potentials.

Comparison of Eqs. (\ref{eq:A-adiab}) and (\ref{eq:A-diab}) for low-energy collisions with 
Eqs. (\ref{eq:1Born}) and (\ref{eq:1Born2}) 
for high-energy collisions shows that the basic physics behind the slow atomic collision theory and 
the first Born approximation, which can be related to the classical description, is different. 
In high-energy collisions the perturbation theory is based on the unperturbed atomic-state wave functions  
and an interaction between a charged impact particle and atomic electrons, 
while in low-energy atomic collisions the perturbation theory is based on the molecular-state wave functions 
and an operator appearing from the nuclear kinetic energy operator. 
If one would use unperturbed electronic atomic-state wave functions within perturbation theory 
for slow atomic collisions, a number of new terms related to other interactions would appear 
in addition to the operators mentioned above.

Thus, it is not surprising that, as we will see, the Drawin formula and the quantum calculations give substantially 
different results. In some cases, simplified nonadiabatic transition models, e.g., 
the Landau-Zener model \citep{Landau:1932a,Landau:1932b,1932RSPSA.137..696Z}, the \citet{Demkov:1964}, 
and the \citet{1962OptSp..13..431N} ones, which are based on the molecular-state representation, can give 
reliable inelastic cross sections. Nevertheless, the most accurate results are obtained 
by means of complete quantum treatments.

\subsection{Quantum collision calculations}
\label{sect:quant}

The basic theory of atomic collisions has been reasonably well understood for some time \citep[e.g.][and references therein]{MottMassey,1962amp..conf..550B}.  However, the detailed calculation of accurate quantum-chemical data for quasi-molecules required as input to the relevant coupled-channel scattering equations, including for excited states, has been a major hurdle.  Further, while the solution of the scattering equations in the standard adiabatic approach appears straightforward, it encounters difficulties such as the electron translation problem \citep{1958RSPSA.245..175B,2001PhRvA..64e2701B,2010PhRvA..82f0701B}.  

Over the last 35 years, our understanding of diatomic hydride molecules and low-energy H atom collisions with atoms, particularly alkalis, has increased markedly.  Detailed theoretical studies of alkali hydride molecular structure began with those by \citet{1975JChPh..62.3367S,1975JChPh..63.2356M,1980JChPh..73.2817O,1985PhRvA..31.3977M}, and these works showed the importance of the ionic configuration, as had been expected from experimental studies \citep[e.g.][]{1936PhRv...50.1028M}.  Studies of ion-pair production and mutual neutralisation processes in alkali-hydrogen collisions followed shortly after \citep{1978PhRvA..17..889J,1985PhRvA..32.3092O,1986JChPh..84..147E,1999JPhB...32.5451D,1999JPhB...32...81C}. 

Though \citet{1986JChPh..84..147E}  briefly looked at excitation of sodium by high-energy H atom collisions, the study of excitation processes came somewhat later.  Motivated by Steenbock and Holweger's work, \citet{1991JPhB...24.4017F} performed an experimental study of $\mathrm{Na}(3s) + \mathrm{H} \rightarrow \mathrm{Na}(3p) + \mathrm{H}$ at low energies (15--1500 eV), though due to experimental difficulties, not down to near the threshold (2.1 eV), which is the relevant regime for the temperatures of interest.  The results were compared with theory by \citet{1992JPhB...25L.101M} at energies above 200 eV.  Revised experimental data, including results down to 10~eV, and detailed quantum scattering calculations down to the threshold, were presented in \citet{1999PhRvA..60.2151B}.  These studies demonstrated, through comparison with the Landau-Zener model and detailed quantum scattering calculations, that the experimental data could be reasonably explained by nonadiabatic transitions associated with avoided crossings, also called pseudocrossings, in the NaH molecular potentials.  At pseudocrossings, adiabatic potentials approach each other and the transition probabilities may become large even for slow collisions, despite that the Massey criterion based atomic energy level spacings predicts the collisions to be nearly adiabatic \citep[e.g.][pg. 582 and 611]{1962amp..conf..550B}.  

\begin{figure}
   \centering
   \includegraphics[width=90mm]{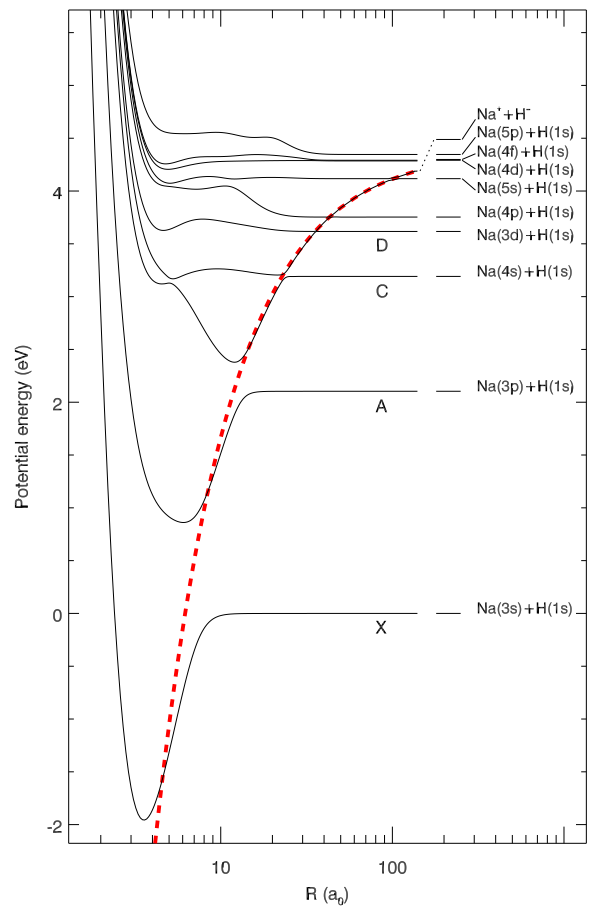}
      \caption{Adiabatic potential energies as a function of internuclear distance $R$ for the lowest ten $^1\Sigma^+$ states of the NaH quasi-molecule from pseudo-potential calculations by \citet{1999JPhB...32.5451D}.  The atomic states and energies at dissociation are shown at the right-hand side of the figure.  The thick, dashed (red) line shows the $1/R$ interaction corresponding to the interaction for the pure ionic configuration Na$^+$+H$^-$ at long range. 
              }
         \label{NaH:pots}
   \end{figure}

In the particular case of low-lying states of NaH, as discussed above, these pseudocrossings arise from interaction of the covalent Na+H and ionic Na$^+$+H$^-$ configurations, and are thus called avoided ionic crossings.  The adiabatic potentials showing these pseudocrossings are shown in Fig.~\ref{NaH:pots}, where we see clearly the influence of the ionic configuration, which strongly perturbs the potentials from their behaviour at large internuclear distance.  Similar effects are seen in other quasi-molecules such as LiH \citep[e.g.][]{1999JPhB...32...81C} and MgH \citep[e.g.][]{guitou2011}.  As shown by diabatic studies, the underlying ionic curve behaves roughly as $1/R$ and pseudocrossings occur with all neutral states below the ionic limit, which show a much weaker dependence on $R$ \citep[e.g.][]{1992JChPh..97.1144B,1995CPL...247...85B}.  Quantum scattering calculations for Li+H \citep{2003PhRvA..68f2703B} and Na+H \citep{2010PhRvA..81c2706B} including all states below the ionic state asymptotic limit have found this to be an important mechanism for nonadiabatic transitions.  Recent calculations for inelastic Mg+H collisions by \citet{guitou2011}, though presently only done for a small number of states, also find this result.  

The basic physical interpretation of this mechanism is that, in a given A+H quasi-molecule, starting from a covalent state with appropriate symmetry where the electrons are associated to their respective atoms, at avoided ionic crossings the valence electron associated with the atom A has a probability to tunnel to the H atom resulting in a predominantly ionic charge distribution A$^+$+H$^-$.  Later during the collision at a different avoided crossing, there is a probability that the electron may tunnel back to a different covalent molecular state leading finally to a different final state of the atom A, and excitation or deexcitation $\mathrm{A}(nl)+\mathrm{H} \rightarrow \mathrm{A}(n'l')+\mathrm{H}$.  The electron may also stay with the H atom leading to ion-pair production $\mathrm{A}+\mathrm{H} \rightarrow \mathrm{A}^++\mathrm{H}^-$.  Calculations for Li+H and Na+H mentioned above have generally shown that large-valued cross sections for excitation are typically small compared with those for ion-pair production from certain states.  This can be easily understood: a large-valued cross section for an excitation process results from passing two avoided ionic crossings, one of them (higher-lying) passed with a small nonadiabatic transition probability, while for an ion-pair production process this region is passed with a large transition probability due to small adiabatic energy splittings at highly excited crossings.

We conclude by emphasizing that the mechanism found to be important in Li+H, Na+H and Mg+H collisions discussed above, that of the avoided ionic crossings, is an example of more general phenomena in atomic collisions, which will be discussed further below.  In He+H inelastic collisions \citet{2001PhRvA..64e2701B} couplings arise due to electronic structure properties completely unrelated to the ionic configuration.  Systems such as Li+Na, on the other hand, show a similar mechanism \citep{2010PhRvA..82f0701B}.

\subsection{Discussion}

Low-energy atomic collisions differ at a fundamental level from electron collisions, especially those at high energies.   This difference arises, naturally, from the presence of a complex projectile and thus the effects of nuclear dynamics in the atomic collisions.  As put succinctly by \citet{1962amp..conf..550B}:  ``They are more complicated in that a wider variety of reaction paths may be followed: for example, the projectile as well as the target may have structure and hence may undergo excitation or ionization; and again, in addition to the possibility of electron exchange, the possibility of charge transfer arises.''   Put another way, the effect of the nuclear dynamics is that the electronic structure of the atom-atom system during the collision can differ significantly from the electronic structure of the separated atoms, leading to interactions and couplings that are very different in character to those found in the separated atoms.    The phenomenon of the avoided ionic crossings, where an electron may tunnel from one nucleus to the other to form an ionic electronic structure leading to strong coupling between adiabatic states, is a particular example of the effects of nuclear dynamics in atomic collisions.  In contrast, electron collisions with atoms have no such possibility to change the structure of the target atom in the same way.  

Thus, we see there is little in common between the physics of the Drawin formula and the physics of low-energy atomic collisions, especially H atom collisions, as we now understand them.   The Drawin formula is based on a classical model and we see that the fundamental mechanism involved in H atom collisions for the cases studied relates to electron tunnelling, which is quantum mechanical in nature.  The Drawin formula extends a theory for electrons, neglecting the atomic nuclei other than in scaling the collision energy.   Thus, from a physics point of view it seems the grounds for expectations that Drawin formula can provide good estimates are rather weak, despite the fact that Thomson formula on which it is based is capable of providing order-of-magnitude estimates in some situations for ionization by electron impact.   

\section{Comparison of results}
\label{sect:comp}

Detailed quantum scattering calculations of cross sections, and corresponding rate coefficients, have been done for excitation processes between all states below the ionic limit for Li+H \citep{2003PhRvA..68f2703B,2003A&A...409L...1B} and Na+H \citep{2010PhRvA..81c2706B,2010A&A...519A..20B}. Recently, calculations for the three lowest states of Mg+H have also been performed \citep{guitou2011}.  In this section we compare the results with the predictions of the Drawin formula.  This comparison will be done on two levels.  Firstly, comparison of the cross sections, which best elucidates the physics.  Secondly, comparison of the rate coefficients, which is of most interest for astrophysical applications, and also allows us to focus on more general integrated properties.

\subsection{Cross sections}

In this section we compare the Drawin cross sections with other theoretical and experimental results. Note, since \citet{1984A&A...130..319S} give only a formula for the rate coefficient, and due to above mentioned problems with Drawin's derivation, the Drawin cross section formula used here is taken from \citet[Eqs. A2 and A8]{1993PhST...47..186L}.  As shown there, these expressions give a formula for the rate coefficient differing from that of \citet{1984A&A...130..319S} by only a factor of $m_\mathrm{A}/(m_\mathrm{A}+m_\mathrm{H})$, which is of order unity. 

\begin{figure}
\centering
\includegraphics[width=90mm]{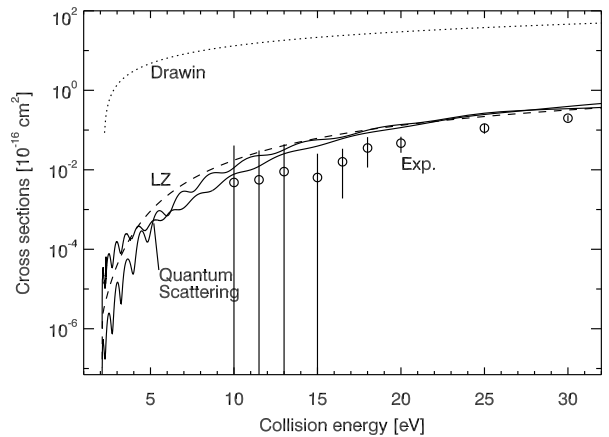}
\caption{Comparison of cross sections for the process Na(3s) + H $\rightarrow$ Na(3p) + H.  The dotted line shows the Drawin cross section and the dashed line the Landau-Zener cross section.  The full lines are two quantum scattering calculations using different input quantum-chemical data (MRDCI, which is largest near the threshold, and pseudopotential; see \citet{2010PhRvA..81c2706B} for details).  The circles show the experimental data of \citet{1991JPhB...24.4017F,1999PhRvA..60.2151B} with 1$\sigma$ error bars.}
         \label{NaH:rescross}
\end{figure}

Fig.~\ref{NaH:rescross} compares cross sections at low energy for $\mathrm{Na}(3s) + \mathrm{H} \rightarrow \mathrm{Na}(3p) + \mathrm{H}$, the single case where experimental data is available.   We see that quantum scattering calculations and the Landau-Zener model results agree quite well with experiment.  Near the threshold, where there is no experimental data, the quantum scattering calculations show substantial differences depending on which quantum-chemical data are used.  This highlights the sensitivity of the near-threshold cross sections to the uncertainties in the quantum-chemistry data, and gives an estimate of the uncertainties in the calculated cross sections.  In this case the uncertainties seem to be around one or two orders of magnitude.  However, it should be noted that the uncertainties vary strongly from transition to transition, and the transitions with the largest cross sections have the smallest uncertainties, as low as a factor of 2.  Fortunately, these are the most important from an astrophysical perspective \citep[see discussion in][]{2010A&A...519A..20B}. 

\begin{figure}
   \centering
   \includegraphics[width=90mm]{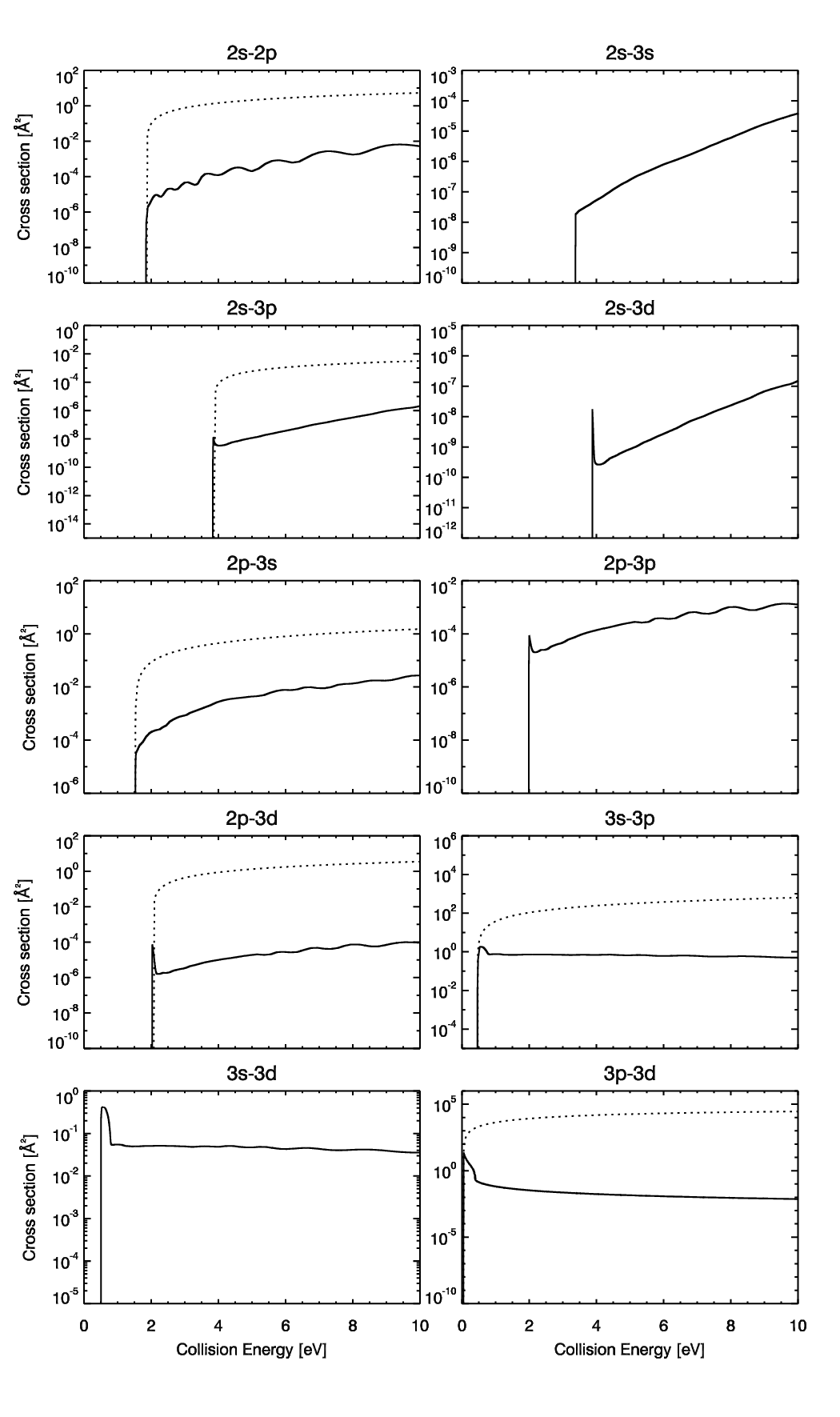}
      \caption{Comparison of quantum scattering cross sections for Li+H from \citet{2003PhRvA..68f2703B} (full lines) with those of the Drawin formula (dotted lines) for the 10 transitions between the 4 lowest states of Li.
              }
         \label{LiH:cross}
   \end{figure}

\begin{figure}
   \centering
   \includegraphics[width=90mm]{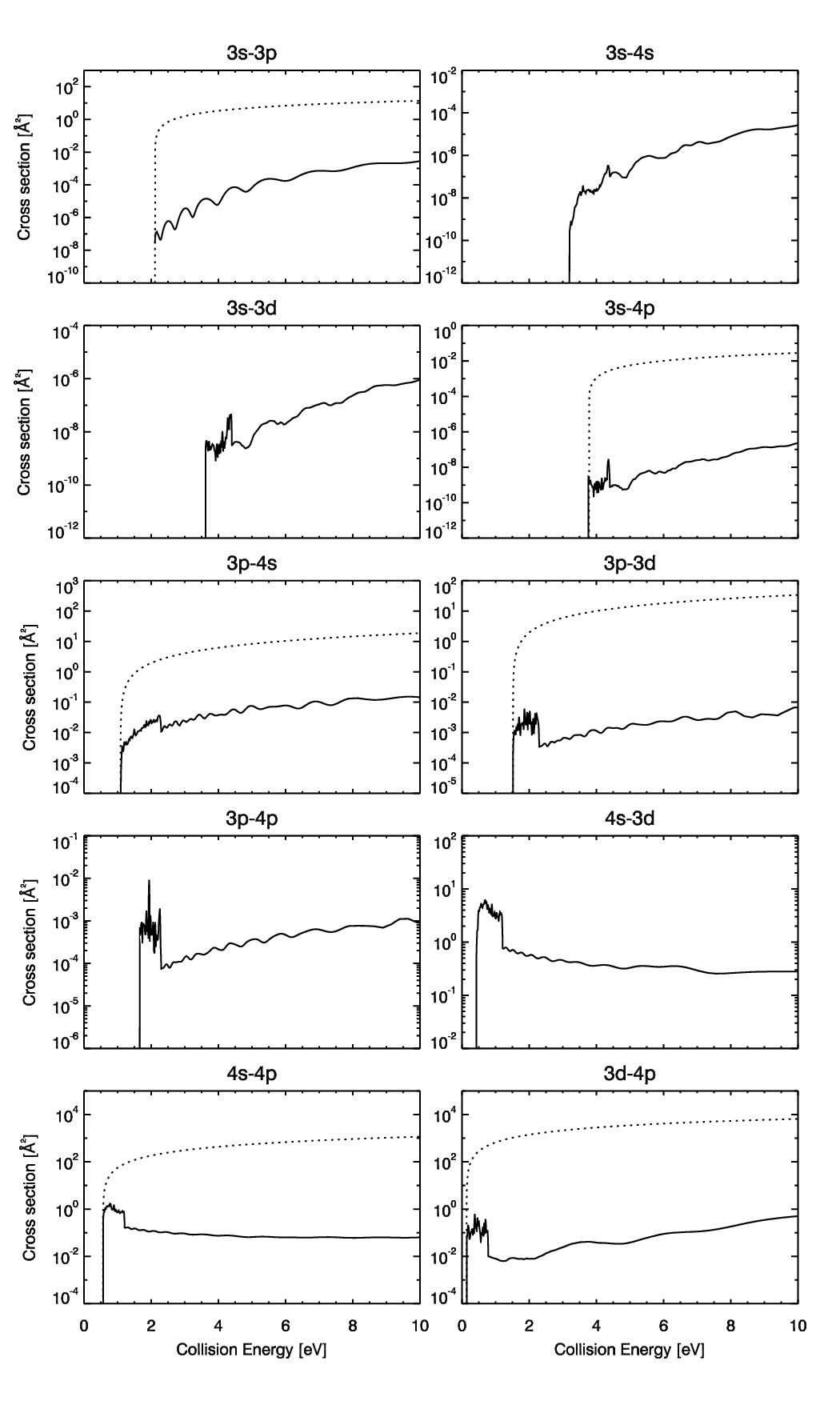}
      \caption{Comparison of quantum scattering cross sections for Na+H from \citet{2010PhRvA..81c2706B} (full lines) with those of the Drawin formula (dotted lines) for the 10 transitions between the 4 lowest states of Na.
              }
         \label{NaH:cross}
   \end{figure}

In Figs.~\ref{LiH:cross} and \ref{NaH:cross}, theoretical excitation cross sections for inelastic Li+H and Na+H collisions at low energy are compared for a number of transitions.    
A major problem of the Drawin formula is immediately evident here: it predicts zero cross sections for optically forbidden transitions (or cross sections much smaller than those for optically allowed transitions if one uses the very small $f$-values in some cases).  The Drawin formula depends on the absorption oscillator strength, while collisional processes do not follow the same selection rules, and no such dependence is seen in the quantum mechanical data.  For the optically allowed transitions the Drawin cross section is typically greater than the quantum scattering results by several orders of magnitude.   The Drawin formula gives a reasonable description of the cross section behaviour with collision energy in many cases, though in a few, particularly among transitions involving excited initial states, the quantum cross sections have a much flatter behaviour with collision energy.

\subsection{Rate coefficients}

\begin{figure*}
   \centering
   \includegraphics[width=90mm]{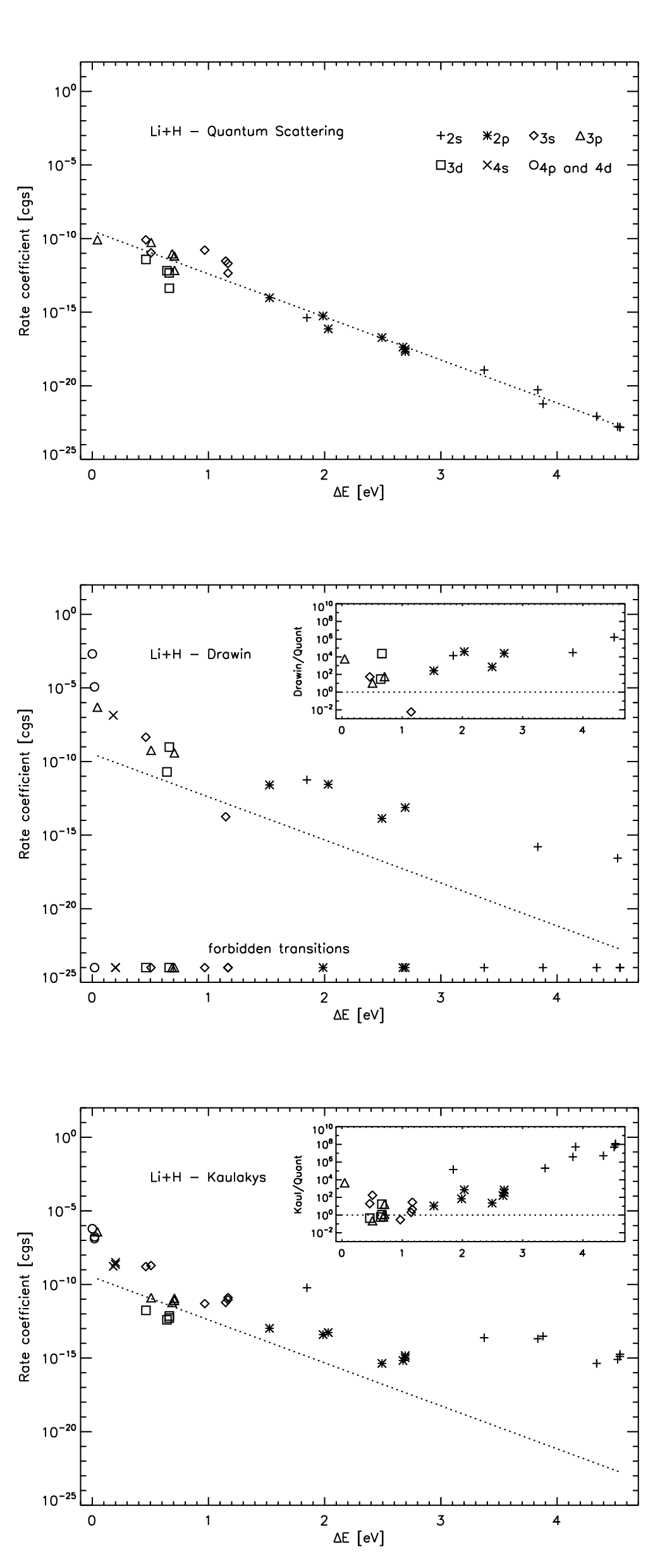}
   \includegraphics[width=90mm]{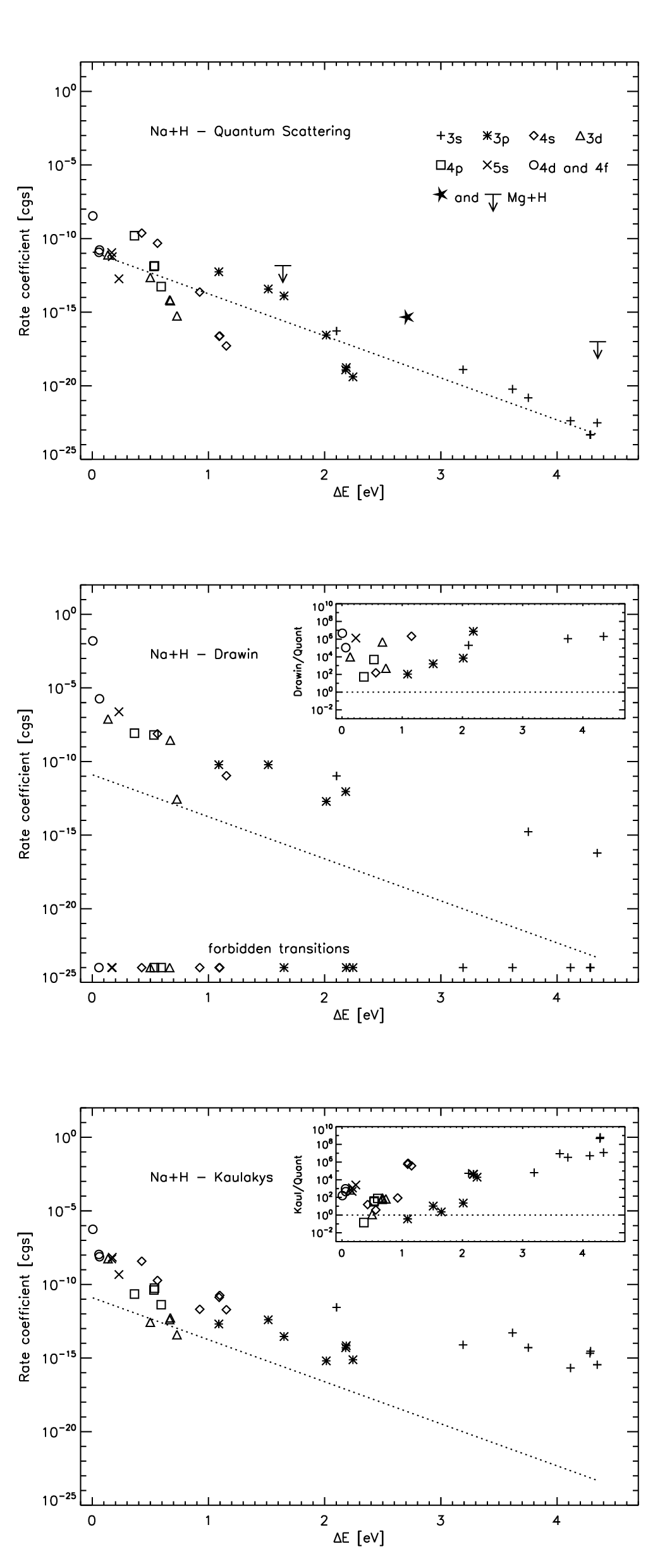}
      \caption{Rate coefficients at 6000~K from various calculations for excitation transitions between the lowest 9 states of Li and Na due to H atom collisions, plotted against the transition energy.  The panels in the left column are for Li+H, and in the right column Na+H.  In each panel, different symbols are used to denote the initial state of the transition following the key given in the upper panels.  The upper panels show the results from detailed quantum scattering calculations \citep{2003A&A...409L...1B, 2010A&A...519A..20B}, noting there are no results for Li $n=4$ initial states as the data for these transitions were deemed unreliable.  The dotted line shows a linear fit to the these data, which is repeated in the lower two panels to aid comparison.  The upper panel in the right column also shows the limited data now available for Mg+H calculated from cross sections given in \citet{guitou2011}.  Note that the data for the two transitions to the $3p \, ^1$P$^o$ state should be interpreted as upper limits; however, they are probably not more than an order of magnitude or two lower.  No data for Mg+H are plotted in the other panels.   The middle panels show the results of the Drawin formula, and the insets show the ratio with the quantum scattering calculations.  Forbidden transitions, where the Drawin formula is not applicable, are shown in the bottom of the panels.  The lower panels show the results of the Kaulakys free electron model \citep{1991JPhB...24L.127K}, and the insets show the ratio with the quantum scattering calculations.
              }
         \label{fig:rates}
   \end{figure*}

In Fig.~\ref{fig:rates}, rate coefficients for excitation at 6000~K by inelastic Li+H, Na+H and Mg+H collisions, found by integration of the cross sections over a Maxwellian velocity distribution, are shown and compared.  In addition to the quantum mechanical calculations and the Drawin formula results, we also calculate rate coefficients from the free electron model described by \citet{1991JPhB...24L.127K}.  Cross sections were calculated according to Eq.~18 in Kaulakys, using non-hydrogenic wave functions in momentum space calculated using the methods of \citet{1997JPhB...30.2403H}.  This model is only valid for transitions in Rydberg atoms, and thus is not valid for the low-lying states being examined here; however, we make a comparison in any case to see if there is any possibility it may give reasonable estimates for transitions among low-lying states.  The rate coefficients are plotted against the transition energy, $\Delta E$, since this a main parameter in determining the rate coefficients, as seen in the figure, due to the strong dependence of the integral on the threshold (e.g., see Eq.~\ref{eq:drawin} where the exponential dependence on $\Delta E=E^\mathrm{A}_{ul}$ is easily seen).  This allows us to separate this effect from other more intrinsic secondary effects related to the physics of the collisions.  The quantum mechanical results correlate strongly with $\Delta E$, though with a significant scatter of several orders of magnitude around the mean relation.  One sees, particularly in the case of Na+H, clear secondary trends among transitions with the same initial state, e.g. $3p$ in Na+H.  These trends result from regular behaviour in the properties of the quasi-molecule at avoided crossings (potential splitting, radial coupling), leading to trends in the transition probabilities.  

Comparison with the Drawin formula immediately shows the problem already discussed: lack of predictions for forbidden transitions.  The Drawin formula data also show a strong correlation with $\Delta E$, as would be expected since for the case of a given element and temperature, the formula has only two parameters, $\Delta E=E^\mathrm{A}_{ul}$ and $f_{lu}$.  Thus, all deviation from the main trend with $\Delta E$ is a result of the $f$-value and those transitions deviating significantly from the trend are those with very high or low $f$-values.  These secondary trends are seen not to correlate with those seen in the quantum mechanical data, which is expected as the physical mechanisms are not related.   In general, it is seen that the rate coefficients differ from the quantum ones by up to eight orders of magnitude with a general tendency to be significantly larger.  There is also a huge scatter spanning around eight orders of magnitude.   
   
Though the Kaulakys free electron model has the advantage of providing data for all transitions, it compares no better with the detailed quantum scattering results in general, except perhaps for transitions with small transition energies $\Delta E$ where agreement is within a few orders of magnitude.  Comparison shows a clear tendency for increasingly poor agreement with increasing transition energy, with significant scatter.  Thus, there is no suggestion that the Kaulakys free electron model could give reasonable estimates for transitions among low-lying states.  It should, however, provide reasonable estimates for transitions among Rydberg states above the ionic limit.

\section{Conclusions}

The Drawin formula does not contain the essential physics behind direct excitation by H atom collisions at low energies, which is quantum mechanical in character.  It compares poorly with the results of the available full quantum scattering calculations based on detailed quantum-chemical modelling of the quasi-molecule. The estimated uncertainties in the rate coefficients from the quantum mechanical data are not sufficient to explain the differences \citep[see, e.g., ][]{2010A&A...519A..20B}.  In general, the Drawin formula significantly overestimates the rates by amounts that vary markedly from transition to transition.   Thus, we conclude that the Drawin formula does not provide order-of-magnitude estimates for excitation by H atom collisions.  We can even more generally conclude that formulae based on the Drawin expressions are also unlikely to be useful for excitation by other atoms.

The Drawin formula is often used in non-LTE stellar atmosphere applications in a semi-empirical fashion with a scaling factor, often denoted $S_\mathrm{H}$.  The problem of forbidden transitions is sometimes avoided with some arbitrary solution to allow data to be estimated, for example an arbitrary constant $f$-value or scaled $f$-value \citep[e.g.][]{1985ASSL..114..231S}.  The value of $S_\mathrm{H}$ is varied either to assess the possible influence of H atom collisions or to calibrate it against some observational criteria.  This procedure, in the absence  of a better alternative, is not unreasonable in view of the fact that $\Delta E$ has a strong influence on the rate coefficient, and this influence resulting from the velocity distribution is accounted for by the Drawin formula.  However, as we have seen, there is a large scatter around this basic behaviour of the rate coefficients, which stems from the physics of the atomic collision, and is not reproduced by the Drawin formula.   This means such a procedure will not give good estimates of relative rates for collisional transitions, and could lead to modelling errors if more than one transition is of importance.  Furthermore, if other atomic collision processes are present but not modelled, effects caused by other processes may be incorrectly attributed to H atom collisions.  For example, in modelling of Li and Na non-LTE line formation it has been shown that the ion-pair production processes mentioned in Sect.~\ref{sect:quant}, often called charge exchange or charge transfer, are far more important than direct excitation \citep{2003A&A...409L...1B,Lind11}.  Deficiencies in other aspects of the spectrum modelling might also be masked by such an approach.

It is clear that there is a need for quantum mechanical estimates for direct excitation and ion-pair production processes.  The most accurate calculations are naturally provided by the full quantum treatment, i.e. detailed quantum-chemical data calculation with detailed quantum scattering calculations.  However, in order to provide estimates for a wide range of elements of astrophysical interest, simplified models, based on the physical understanding provided by the detailed calculations, would be required.  The reasonable success of the Landau-Zener model, as seen in Fig.~\ref{NaH:rescross}, suggests this model, when coupled with a method for producing quantum-chemical data, could lie at the basis of attempts to produce such simplified models.  However, we emphasise that more experience from the full quantum calculations is complementary and still needed.

\begin{acknowledgements}
We acknowledge the important contributions of Alan Dickinson, who sadly passed away last year, to parts of this work.  We gratefully acknowledge the support of the Royal Swedish Academy of Sciences, the Wenner-Gren foundation, G{\"o}ran Gustafssons Stiftelse and the Swedish Research Council.  P.S.B is a Royal Swedish Academy of Sciences Research Fellow supported by a grant from the Knut and Alice Wallenberg Foundation.   P.S.B is grateful for hospitality of the Max Planck Institute for Astrophysics, Garching, where parts of this article were written.  A.K.B also acknowledges support from the Russian Foundation for Basic Research (grant No.\ 10-03-00807-a).  The authors acknowledge the role of the SAM collaboration (http://www.anst.uu.se/ulhei450/GaiaSAM/ ) in stimulating this research through regular workshops.
\end{acknowledgements}

\bibliographystyle{aa} 
\bibliography{16745} 

\end{document}